\documentclass[preprint,aps,12pt,showpacs,nofootinbib,tightenlines]{revtex4}
\usepackage{mathrsfs}
\usepackage{amsmath}
\usepackage{amssymb}
\usepackage{epsfig}
\usepackage{graphicx}
\newcommand{\psl}{ P \hspace{-2.4truemm}/ }
\newcommand{\nsl}{ n \hspace{-2.2truemm}/ }
\newcommand{\vsl}{ v \hspace{-2.2truemm}/ }
\newcommand{\esl}{ \epsilon \hspace{-1.5truemm}/ }

\def\be{\begin{eqnarray}}
\def\en{\end{eqnarray}}
\def\non{\nonumber\\}

\def\prd{{Phys. Rev. D}~}

\def\plb{{ Phys. Lett. B}~}

\newcommand{\acp}{{\cal A}_{CP}}
\begin{document}
\title{$B\to K_1\pi(K)$ decays in the perturbative QCD approach}
\author{Zhi-Qing Zhang$^1$
,~ Zhi-Wei Hou$^1$,~ Yueling Yang$^2$,~ Junfeng Sun$^2$
\\
{\small $^1$ \it Department of Physics, Henan University of
Technology,}\\
{\small \it Zhengzhou, Henan 450001, China;} \\
{\small $^2$ \it College of Physics and Information Engineering,}\\
{\small \it Henan Normal University, Xinxiang 453007, China}}
\date{\today}
\begin{abstract}
Within the framework of the perturbative QCD approach, we study the
two-body charmless decays $B\to K_1(1270)(K_1(1400))\pi(K)$. We find
the following results: (i) The decays $\bar B^0\to K_1(1270)^+\pi^-, K_1(1400)^+\pi^-$ are
incompatible with the present experimental data. There exists a similar situation for the decays $\bar B^0\to a_1(1260)^+K^-, b_1(1235)^+K^-$, which are usually
considered that the nonperturbative contributions are needed to explain the data. But the difference is that the nonperturbative
contributions seem to play opposite roles in these two groups of decays.(ii) The pure annihilation type decays $\bar B^0\to K_1^{\pm}(1270)K^{\mp}, K_1^{\pm}(1400)K^{\mp}$
are good channels to test whether an approach can be used to calculate correctly the strength of the
penguin-annihilation amplitudes. Their branching ratios are predicted at $10^{-7}$ order, which are larger than the QCDF results. (iii)
The dependence of the direct CP-violating asymmetries of these decays on the mixing angle $\theta_{K_1}$ are also considered.
\end{abstract}

\pacs{13.25.Hw, 12.38.Bx, 14.40.Nd}
\vspace{1cm}

\maketitle


\section{Introduction}\label{intro}
In general, the mesons are classified in $J^{PC}$ multiplets. There
are two types of orbitally excited axial-vector mesons, namely
$1^{++}$ and $1^{+-}$. The former includes $a_1(1260), f_1(1285),
f_1(1420)$ and $K_{1A}$, which compose the $^3P_1$-nonet, and the
latter includes $b_1(1235), h_{1}(1170), h_1(1380)$ and $K_{1B}$,
which compose the $^1P_1$-nonet. Except $a_1(1260)$ and $b_1(1235)$,
other axial-vector mesons exist mixing problem, which makes their
inner structure become more ambiguous, for example, $K_{1A}$ and
$K_{1B}$ can mix with each other and form two physical mass
eigenstates $K_1(1270), K_1(1400)$. Various values about the mixing
angle $\theta_{K_1}$ can be found in different literatures, which
will be examined in more detail in Sec.\ref{results}. For the mixings of the
SU(3)-singlet and SU(3)-octet mesons, specifically,  the
$f_1(1285)-f_1(1420)$ mixing angle $\theta_{^3P_1}$ and the
$h_1(1170)-h_1(1380)$ mixing angle $\theta_{^1P_1}$, there also exist several values in the phenomenal analysis. Certainly,
these two angles can associate with $\theta_{K_1}$ through the
Gell-Mann-Okubo mass formula. For the lack of sufficient experimental
data, none of them can be accurately determined up to now. So the
decays involving these mesons become more ambiguous compared with
the decays involving $a_1(1260)$ or/and $b_1(1235)$ meson(s), which
have been discussed in the previous works \cite{wwang,zqzhang1,zqzhang2,cmv,vnp,cy}.

In this paper, we would like to discuss the decays $B\to K_1(1270)\pi(K), K_1(1400)\pi(K)$. On the theoretical side, many approaches have been
used to study these decays, such as the naive factorization \cite{cmv}, the generalized factorization \cite{vnp}, and the QCD factorization approach \cite{cy}.
From the predictions
of these approaches, One can find that the branching ratios of the decays $B\to K_1(1270)\pi,K_1(1400)\pi$ are in the order of $10^{-6}$, for example,
$Br(B^0\to K_1(1270)^+\pi^-)=(3\sim 8)\times10^{-6}$, $Br(B^0\to K_1(1400)^+\pi^-)=(2\sim 5)\times10^{-6}$, those of almost all the decays
$B\to K_1(1270)K,K_1(1400)K$  are in the order of $10^{-8}\sim 10^{-7}$. While on the experimental side, the large upper
limits are given for the decays $B^0\to K_1(1400)^+\pi^-$ and $B^+\to K_1(1400)^0\pi^+$ at the $90\%$ level (C.L.) of $1.1\times10^{-3}$ and $2.6\times10^{-3}$,
respectively \cite{argus}, and the Heavy Flavor Averaging Group(HFAG) gives the following results \cite{hfag}:
\be
Br(B^+\to K_1(1270)^0\pi^+)&<&40\times 10^{-6}, Br(B^+\to K_1(1270)^0\pi^+)<39\times 10^{-6},\label{data1}\\
Br(B^0\to K_1(1270)^+\pi^-)&=&(17^{+8}_{-11})\times 10^{-6}, Br(B^0\to K_1(1400)^+\pi^-)=(17^{+7}_{-9})\times 10^{-6}.\label{data2}\;\;\;\;\;
\en
The preliminary data are given by BABAR \cite{barbar1},
\be
BR(B^0\to K_1^+(1270)\pi^-)&=&(12.0\pm3.1^{+9.3}_{-4.5})\times10^{-6}, \label{data3}\\
BR(B^0\to K_1^+(1400)\pi^-)&=&(16.7\pm2.6^{+3.5}_{-5.0})\times10^{-6}. \label{data4}
\en
Furthermore, BABAR has also measured the branching ratios $Br(B^0\to K_1(1270)^+\pi^-+K_1(1400)^+\pi^-)=3.1^{+0.8}_{-0.7}\times10^{-5}$
and $Br(B^+\to K_1(1270)^0\pi^++K_1(1400)^0\pi^+)=2.9^{+2.9}_{-1.7}\times10^{-5}$ with $7.5\sigma$ and $3.2\sigma$ significance, respectively.
In the paper \cite{barbar2}, the two sided intervals for
some of the decays $B\to K_1(1270)\pi, K_1(1400)\pi$ are evaluated at $68\%$ probability ($\times 10^{-5}$):
\be
BR(B^-\to \bar K_1(1270)^0\pi^-) &\in & [0.0,2.1], BR(B^-\to \bar K_1(1400)^0\pi^-) \in [0.0,2.5],\label{data5}\\
BR(B^0\to K_1(1270)^+\pi^-)&\in & [0.6,2.5], BR(B^0\to K_1(1400)^+\pi^-)\in [0.8,2.4]. \label{data6}\en

In view of the differences between the theories and experiments, we are going to use the PQCD approach to explore these decays and analyze
whether the nonperturbtive contributions
are necessary to explain the experimental data.
In the following, $K_1(1270)$ and $K_1(1400)$ are denoted as $K_1$ in some places for
convenience. The layout of this paper is as follows. In Sec.\ref{proper}, the decay constants and the light-cone
distribution amplitudes of the relevant mesons are introduced. In Sec.\ref{results}, we then analyze these decay channels by using the PQCD
approach. The numerical results and the discussions are given in
Sec. \ref{numer}. The conclusions are presented in the final part.


\section{decay constants and distribution amplitudes }\label{proper}

For the wave function of the heavy $B$ meson,
we take
\be
\Phi_{B}(x,b)=
\frac{1}{\sqrt{2N_c}} (\psl_{B} +m_{B}) \gamma_5 \phi_{B} (x,b).
\label{bmeson}
\en
Here only the contribution of Lorentz structure $\phi_{B} (x,b)$ is taken into account, since the contribution
of the second Lorentz structure $\bar \phi_{B}$ is numerically small \cite{cdlu} and has been neglected. For the
distribution amplitude $\phi_{B}(x,b)$ in Eq.(\ref{bmeson}), we adopt the following model:
\be
\phi_{B}(x,b)=N_{B}x^2(1-x)^2\exp[-\frac{M^2_{B}x^2}{2\omega^2_b}-\frac{1}{2}(\omega_bb)^2],
\en
where $\omega_b$ is a free parameter, we take $\omega_b=0.4\pm0.04$ Gev in numerical calculations, and $N_B=101.4$
is the normalization factor for $\omega_b=0.4$.

The distribution amplitudes of the axial-vector $K_1$ are written as
: \be \langle K_1(P, \epsilon^*_L)|\bar
q_{2\beta}(z)q_{1\alpha}(0)|0\rangle&=&\frac{i\gamma_5}{\sqrt{2N_c}}\int^1_0dx
\; e^{ixp\cdot z}[m_{K_1}\esl^*_L\phi_{K_1}(x)+\esl^*_L \psl\phi_{K_1}^{t}(x)
+m_{K_1}\phi^{s}_{K_1}(x)]_{\alpha\beta},\non \langle K_1(P,
\epsilon^*_T)|\bar
q_{2\beta}(z)q_{1\alpha}(0)|0\rangle&=&\frac{i\gamma_5}{\sqrt{2N_c}}\int^1_0dx
\; e^{ixp\cdot z}\left[m_{K_1}\esl^*_T\phi^v_{K_1}(x)+\esl^*_T
\psl\phi_{K_1}(x) \right.\non &&
\left.+m_{K_1}i\epsilon_{\mu\nu\rho\sigma}\gamma_5\gamma^\mu\epsilon^{*v}_Tn^\rho
v^\sigma\phi^{a}_{K_1}(x)\right]_{\alpha\beta}, \en
where $K_1$ refers to the two flavor states $K_{1A}$ and $K_{1B}$, and the corresponding distribution functions can
be calculated by using light-cone QCD sum rule and listed as follows:
\be
\begin{cases}
\phi_{K_1}(x)=\frac{f_{K_1}}{2\sqrt{2N_c}}\phi_\parallel(x), \phi^T_{K_1}(x)=\frac{f_{K_1}}{2\sqrt{2N_c}}\phi_\perp(x),\\
\phi^t_{K_1}(x)=\frac{f_{K_1}}{2\sqrt{2N_c}}h^{(t)}_\parallel(x), \phi^s_{K_1}(x)=\frac{f_{K_1}}{2\sqrt{4N_c}}\frac{d}{dx}h^{(s)}_\parallel(x),\\
\phi^v_{K_1}(x)=\frac{f_{K_1}}{2\sqrt{2N_c}}g^{(v)}_\perp(x), \phi^a_{K_1}(x)=\frac{f_{K_1}}{8\sqrt{2N_c}}\frac{d}{dx}g^{(a)}_\perp(x). \label{vamp}
\end{cases}
\en
Here we use $f_{K_{1}}$ to present both the longitudinally and transversely polarized states $K_{1A}(K_{1B})$
by assuming $f^T_{K_{1A}}=f_{K_{1A}}=f_{K_1}$ for $K_{1A}$
and $f_{K_{1B}}=f^T_{K_{1B}}=f_{K_1}$ for $K_{1B}$, respectively.
It is similar for the case of $a_1 (b_1)$ states, and the difference is that here $K_{1A}$ and $K_{1B}$ are not the mass eigenstates.
In Eq.(\ref{vamp}), the twist-2 distribution functions are in the first line and can be expanded as:
\be
\phi_{\parallel,\perp}&=&6x(1-x)\left[a^{\parallel,\perp}_0+3a^{\parallel,\perp}_1t+a^{\parallel,\perp}_2\frac{3}{2}(5t^2-1)\right],
\en
the twist-3 light-cone distribution amplitudes (LCDAs) are used the following forms for $K_{1A}$ and $K_{1B}$ states:
\be
h^{(t)}_\parallel(x)&=&3a^\perp_0t^2+\frac{3}{2}a^\perp_1t(3t^2-1), h^{(s)}_\parallel(x)=6x(1-x)(a^\perp_0+a^\perp_1t),\non
g^{(a)}_\perp(x)&=&6x(1-x)(a^\parallel_0+a^\parallel_1t), g^{(v)}_\perp(x)=\frac{3}{4}a^\parallel_0(1+t^2)+\frac{3}{2}a^\parallel_1t^3, \label{t4}
\en
where $t=2x-1$ and the Gegenbauer moments \cite{cheng} $a^{\perp}_0(K_{1A})=0.26^{+0.03}_{-0.22}, a^{\parallel}_0(K_{1B})=-0.15\pm0.15,
a^{\parallel}_0(K_{1A})=a^{\perp}_0(K_{1B})=1$,
$a^{\perp}_1(K_{1A})=-1.08\pm0.48, a^{\perp}_1(K_{1B})=0.30^{+0.00}_{-0.31}$, $a^{\parallel}_1(K_{1A})=-0.30^{+0.26}_{-0.00}$
, $a^{\parallel}_1(K_{1B})=-1.95\pm0.45$, $a^{\parallel}_2(K_{1A})=-0.05\pm0.03, a^{\parallel}_2(K_{1B})=0.09^{+0.16}_{-0.18}$.

The wave functions for the pseudoscalar (P) mesons $K, \pi$ are
given as: \be \Phi_{K(\pi)}(P,x,\zeta)\equiv
\frac{1}{\sqrt{2N_C}}\gamma_5 \left [ \psl \phi^{A}_{K(\pi)}(x)+m_0
\phi^{P}_{K(\pi)}(x)+\zeta m_0 (\vsl \nsl - v\cdot
n)\phi^{T}_{K(\pi)}(x)\right ],  \en
where the parameter $\zeta$ is either $+1$ or $-1$ depending on the
assignment of the momentum fraction $x$. The chiral scale parameter $m_0$ is defined as $m_0=\frac{m^2_\pi}{m_{u}+m_{d}}$ for $\pi$ meson and
$m_0=\frac{m^2_K}{m_{u}+m_{s}}$ for $K$ meson.
The distribution amplitudes are expanded as:
\be
\phi^A_{K(\pi)}(x)&=&\frac{3f_{K(\pi)}}{\sqrt{6}}x(1-x)\left[1+a_{1K(\pi)}C^{3/2}_1(t)+a_{2K(\pi)}C^{3/2}_2(t)+a_{4K(\pi)}C^{3/2}_4(t)\right],\label{kpi}\\
\phi^P_{K(\pi)}(x)&=&\frac{3f_{K(\pi)}}{2\sqrt{6}}\left[1+\left(30\eta_3-\frac{5\rho^2_{K(\pi)}}{2}\right)C^{1/2}_{2}(t)\right.\non
&&\left.-3\left( \eta_3\omega_3+\frac{9\rho^2_{K(\pi)}}{20}(1+6a_{2K(\pi)})\right)C^{1/2}_4(t)\right],\\
\phi^T_{K(\pi)}(x)&=&\frac{-f_{K(\pi)}t}{2\sqrt{6}}\left[1+6(5\eta_3-\frac{\eta_3\omega_3}{2}-\frac{7\rho^2_{K(\pi)}}{20}-\frac{3\rho^2_{K(\pi)}a_{2K(\pi)}}{5})
(1-10x+10x^2)\right],\;\;\;\;\;
\en
where the decay constants $f_{K}=0.16$ GeV, $f_{\pi}=0.13$ GeV and the Gegenbauer moments, Gegenbauer polynomials are defined as:
\be
a_{1K}&=&0.17\pm0.17, a_{1\pi}=0, a_{2K}=a_{2\pi}=0.115\pm0.115, a_{4K}=a_{4\pi}=-0.015,\non
C^{3/2}_1(t)&=&3t, C^{3/2}_2(t)=\frac{3}{2}(5t^2-1), C^{3/2}_{4}(t)=\frac{15}{8}(1-14t^2+21t^4),\non
C^{1/2}_2(t)&=&\frac{1}{2}(3t^2-1), C^{1/2}_{4}(t)=\frac{1}{8}(3-30t^2+35t^4),  \label{kpi1}
\en
and the constants $\eta_3=0.015,\omega_3=-3$,
the mass ratio $\rho_{K(\pi)}=m_{K(\pi)}/m_{0K(\pi)}$ with $m_K=0.49$ GeV, $m_{0K}=1.7$ GeV, $m_\pi=0.135$ GeV, $m_{0\pi}=1.4$ GeV.

\section{ the perturbative QCD  calculation} \label{results}
The PQCD approach is an effective theory to handle hadronic $B$ decays \cite{cdlu2,keum,mishima}.
Because it takes into account the transverse momentum of the valence
quarks in the hadrons, one will encounter the double logarithm divergences when the soft and the collinear momenta overlap. Fortunately, these large
double logarithm can be resummed into the Sudakov factor \cite{hnli0}. There also exit another type of double logarithms which arise from the loop corrections
to the weak decay vertex. These double logarithms can also be resummed and resulted in the threshold factor \cite{hnli00}. This factor decreases faster than any other
power of the momentum fraction in the threshold region, which removes the endpoint singularity. It is often parameterized
into a simple form which is independent on channels, twists and flavors \cite{hnli}. Certainly, when the higher order diagrams only suffer from soft or
collinear infrared divergence, it is ease to cure by using the eikonal approximation \cite{hnli2}. Controlling these
kinds of divergences reasonably makes the PQCD approach more self-consistent.
\begin{figure}[t,b]
\vspace{-3cm} \centerline{\epsfxsize=16 cm \epsffile{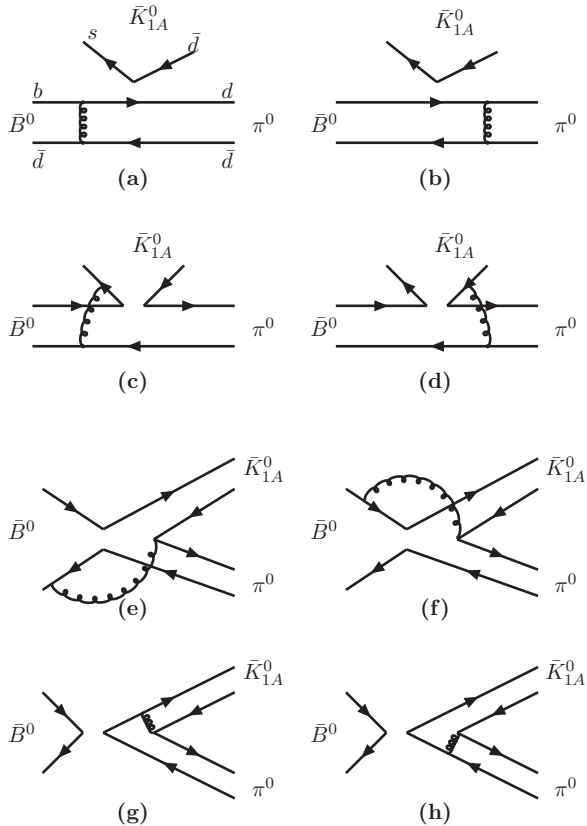}}
\vspace{-9cm} \caption{ Diagrams contributing to the decay $\bar{B}^0\to \bar K^0_{1A}\pi^0$.}
 \label{fig1}
\end{figure}

For these two axial vector mesons, their mass eigenstates and
flavor eigenstates are not the same with each other, and the former can be
obtained by the latter through a mixing angle
$\theta_{K_1}$: \be
K_1(1270)=K_{1A}\sin\theta_{K_1}+K_{1B}\cos\theta_{K_1},
K_1(1400)=K_{1A}\cos\theta_{K_1}-K_{1B}\sin\theta_{K_1}. \label{thetak1} \en
Unfortunately, there are many uncertainties about this mixing
angle. From various phenomenological analysis and experimental data on the masses of these two physical states, it indicates
that this mixing angle is around either $33^\circ$ or $58^\circ$ \cite{rkc,iw,dma,su,bg,pvc,gi,vfv,tky,div}.
Certainly, the author of \cite{cheng1} stresses that the sign of $\theta_{K_1}$ depends on the relative sign of flavor states $K_{1A}$
and $K_{1B}$, which can be determined by fixing the relative sign of the decay constants of $K_{1A}$ and $K_{1B}$. If the decay
constants $f_{1A}, f_{1B}$ are the same in sign (it means that the transitions $B\to K_{1A}$ and $B\to K_{1B}$ have the opposite signs), then
the mixing angle $\theta_{K_1}$ defined in (\ref{thetak1}) is positive. It is noticed that the mixing angle for the antiparticle states
$\bar K_{1}(1270), \bar K_{1}(1400)$, which is denoted as $\theta_{\bar K_1}$, is of opposite sign to that for the particle states $K_{1}(1270), K_1(1400)$.
But even so, we cannot confirm whether $\theta_{K_1}$ is larger or less than $45^\circ$ up to now. Different approaches and models
are used and different values of the mixing angle are obtained. In order to pin down it, Cheng \cite{cheng1} advocates to determine the mixing
angles $\theta_{^3P_1}$ and $\theta_{^1P_1}$ between $f_1(1285)-f_1(1420)$ and $h_1(1170)-h_1(1380)$, respectively,  which in turn depend on the
$K_{1A}-K_{1B}$ mixing angle $\theta_{K_1}$ through the mass relation. Through analyzing the present data of the $h_1, f_1$ mesons' strong/radiative
decay modes, the author prefers $\theta_{K_1}\sim 33^\circ$ over $58^\circ$. In view of the present limited data, we will still include the mixing angle
$\theta_{K_1}\sim 58^\circ$ in our calculations.

It is just because of the ambiguous mixing angle that makes the study very difficult. Here we take the decay
$\bar B^0\to \bar K_1(1270)^0\pi^0$ as an example, which is contributed by the decays $\bar B^0\to \bar K^0_{1A}\pi^0$ and
$\bar B^0\to \bar K^0_{1B}\pi^0$. Figure 1 is for the Feynman diagrams of the decay $\bar B^0\to \bar K^0_{1A}\pi^0$ (it is similar to the decay
$\bar B^0\to \bar K^0_{1B}\pi^0$), through which the amplitudes can be calculated directly,
and the total amplitudes of the decay $\bar B^0\to \bar K_1(1270)^0\pi^0$ can be obtained by combining the two sets of
flavor state amplitudes according to Eq.(\ref{thetak1}):
\be
\sqrt 2A(\bar K_1(1270)^{0}\pi^0)&=&-\xi_t(f_{K_{1A}}\sin\theta_{K_1}+f_{K_{1B}}\cos\theta_{K_1})F^{LL}_{e\pi}(a_4-\frac{1}{2}a_{10})\non &&
-\xi_t(M^{LL;K_{1A}}_{e\pi}\sin\theta_{K_1}+M^{LL;K_{1B}}_{e\pi}\cos\theta_{K_1})(C_3-\frac{1}{2}C_9)\non &&
-\xi_t(M^{LR;K_{1A}}_{e\pi}\sin\theta_{K_1}+M^{LR;K_{1B}}_{e\pi}\cos\theta_{K_1})(C_5-\frac{1}{2}C_7)\non &&
-\xi_t(M^{LL;K_{1A}}_{a\pi}\sin\theta_{K_1}+M^{LL;K_{1B}}_{a\pi}\cos\theta_{K_1})(C_3-\frac{1}{2}C_9)\non &&
-\xi_t(M^{LR;K_{1A}}_{a\pi}\sin\theta_{K_1}+M^{LR;K_{1B}}_{a\pi}\cos\theta_{K_1})(C_5-\frac{1}{2}C_7)\non &&
-\xi_tf_B(F^{LL;K_{1A}}_{a\pi}\sin\theta_{K_1}+F^{LL;K_{1B}}_{a\pi}\cos\theta_{K_1})(a_4-\frac{1}{2}a_{10})\non &&
-\xi_tf_B(F^{SP;K_{1A}}_{a\pi}\sin\theta_{K_1}+F^{SP;K_{1B}}_{a\pi}\cos\theta_{K_1})(a_6-\frac{1}{2}a_8)\non &&
+f_{\pi}(F^{LL}_{eK_{1A}}\sin\theta_{K_1}+F^{LL}_{eK_{1B}}\cos\theta_{K_1})\left[\xi_ua_1
-\xi_t\left(\frac{3C_9}{2}+\frac{C_{10}}{2}\right.\right.\non && \left.\left.-\frac{3C_7}{2}-\frac{C_8}{2}\right)\right]
+(M^{LL;\pi}_{eK_{1A}}\sin\theta_{K_1}+M^{LL;\pi}_{eK_{1B}}\cos\theta_{K_1})\left[\xi_uC_2\right.\non &&\left.-\xi_t\frac{3C_{10}}{2}\right]
-\xi_t (M^{SP;\pi}_{eK_{1A}}\sin\theta_{K_1}+M^{SP;\pi}_{eK_{1B}}\cos\theta_{K_1})\frac{3C_8}{2}, \label{kpi1}
\en
where $\xi_u=V_{ub}V^*_{us}, \xi_t=V_{tb}V^*_{ts}$, $F^{M_2}_{e(a)M_1}$ and $M^{M_2}_{e(a)M_1}$ denote the amplitudes of factorizable and nonfactorizable emission (annihilation)
diagrams, where the subscript meson $M_1$ is involved in the $\bar B^0$ meson transition, the superscript meson $M_2$ is the emitted particle. The other superscript in each amplitude
denotes different current operators, $(V-A)(V-A), (V-A)(V+A)$ and $(S-P)(S+P)$ corresponding to $LL, LR$ and $SP$, respectively. If exchanging the
positions of $K_{1A}$ and $\pi^0$ in Fig.1(a), 1(b), 1(c) and 1(d),  we will get the new Feynman diagrams, which can also contribute to the decay
$\bar B^0\to \bar K^0_{1A}\pi^0$, and the corresponding amplitudes are given in the last three lines of Eq.(\ref{kpi1}). The amplitudes for
the decay $ \bar B^0\to \bar K^0_{1A}(\bar K^0_{1B})\pi^0$ can be obtained from those for the decay $B\to K\pi$ which can be found in \cite{ali},
only changing the variables of $K$ meson with those
of $K^0_{1A}(K^0_{1B})$ meson. So we do not list the analytic expressions for these amplitudes. Certainly, it is noticed that if the axial-vector meson $K_{1A}(K_{1B})$ is on the emitted
position in the factorizable emission diagrams, there is no scalar or pseudoscalar current contribution.
The total amplitudes for the other three $B\to K_1(1270)\pi$ decay modes can also be written out similarly:
\be
A(K_1(1270)^{-} \pi^+)&=&(f_{K_{1A}}\sin\theta_{K_1}+f_{K_{1B}}\cos\theta_{K_1})F^{LL}_{e\pi}(\xi_u a_1
-\xi_t(a_4+a_{10})) \non  &&+(M^{LL;K_{1A}}_{e\pi}\sin\theta_{K_1}+M^{LL;K_{1B}}_{e\pi}\cos\theta_{K_1})(\xi_u C_1-\xi_t(C_3+C_9))
\non   &&-\xi_t
(M^{LR;K_{1A}}_{e\pi}\sin\theta_{K_1}+M^{LR;K_{1B}}_{e\pi}\cos\theta_{K_1})(C_5+C_7)\non   &&-
\xi_t(M^{LL;K_{1A}}_{a\pi}\sin\theta_{K_1}+M^{LL;K_{1B}}_{a\pi}\cos\theta_{K_1})(C_3-\frac{1}{2}C_9) \non  &&
-\xi_t(M^{LR;K_{1A}}_{a\pi}\sin\theta_{K_1}+M^{LR;K_{1A}}_{a\pi}\cos\theta_{K_1})(C_5-\frac{1}{2}C_7)\non  &&-\xi_tf_B
(F^{LL;K_{1A}}_{a\pi}\sin\theta_{K_1}+F^{LL;K_{1B}}_{a\pi}\cos\theta_{K_1})(a_4-\frac{1}{2}a_{10})  \non &&
-\xi_tf_B(F^{SP;K_{1A}}_{a\pi}\sin\theta_{K_1}+F^{SP;K_{1B}}_{a\pi}\cos\theta_{K_1})(a_6-\frac{1}{2}a_8),\label{kpi2}\\
\sqrt 2A(K_1(1270)^{-}\pi^0)&=&(f_{K_{1A}}\sin\theta_{K_1}+f_{K_{1B}}\cos\theta_{K_1})F^{LL}_{e\pi}\left[\xi_u a_1
-\xi_t(a_4+a_{10})\right]\non &&+(M^{LL;K_{1A}}_{e\pi}\sin\theta_{K_1}+M^{LL;K_{1B}}_{e\pi}\cos\theta_{K_1})\left[\xi_u C_1-\xi_t(C_3+C_9)\right]
\non &&-\xi_t (M^{LR;K_{1A}}_{e\pi}\sin\theta_{K_1}+M^{LR;K_{1B}}_{e\pi}\cos\theta_{K_1})(C_5+C_7)\non   &&
+(M^{LL;K_{1A}}_{a\pi}\sin\theta_{K_1}+M^{LL;K_{1B}}_{a\pi}\cos\theta_{K_1})\left[\xi_uC_1-\xi_t(C_3+C_9)\right] \non &&
-\xi_t(M^{LL;K_{1A}}_{a\pi}\sin\theta_{K_1}+M^{LL;K_{1B}}_{a\pi}\cos\theta_{K_1})(C_5+C_7)\non &&
+f_B(F^{LL;K_{1A}}_{a\pi}\sin\theta_{K_1}+F^{LL;K_{1B}}_{a\pi}\cos\theta_{K_1})\left[\xi_u a_2-\xi_t(a_4+a_{10})\right]\non &&
-f_B(F^{SP;K_{1A}}_{a\pi}\sin\theta_{K_1}+F^{SP;K_{1B}}_{a\pi}\cos\theta_{K_1})\xi_t(a_6+a_8)\non &&
+f_{\pi}(F^{LL}_{eK_{1A}}\sin\theta_{K_1}+F^{LL}_{eK_{1B}}\cos\theta_{K_1})\left[\xi_ua_1
-\xi_t\left(\frac{3C_9}{2}+\frac{C_{10}}{2}\right.\right.\non && \left.\left.-\frac{3C_7}{2}-\frac{C_8}{2}\right)\right]
+(M^{LL;\pi}_{eK_{1A}}\sin\theta_{K_1}+M^{LL;\pi}_{eK_{1B}}\cos\theta_{K_1})\left[\xi_uC_2\right.\non &&\left.-\xi_t\frac{3C_{10}}{2}\right]
-\xi_t (M^{SP;\pi}_{eK_{1A}}\sin\theta_{K_1}+M^{SP;\pi}_{eK_{1B}}\cos\theta_{K_1})\frac{3C_8}{2}, \label{kpi3}
\en
\be
A(\bar K_1(1270)^{0}\pi^-)&=&-\xi_t(f_{K_{1A}}\sin\theta_{K_1}+f_{K_{1B}}\cos\theta_{K_1})F^{LL}_{e\pi}(a_4-\frac{1}{2}a_{10})\non &&
-\xi_t(M^{LL;K_{1A}}_{e\pi}\sin\theta_{K_1}+M^{LL;K_{1B}}_{e\pi}\cos\theta_{K_1})(C_3-\frac{1}{2}C_9)\non &&
-\xi_t(M^{LR;K_{1A}}_{e\pi}\sin\theta_{K_1}+M^{LR;K_{1B}}_{e\pi}\cos\theta_{K_1})(C_5-\frac{1}{2}C_7)\non &&
+(M^{LL;K_{1A}}_{e\pi}\sin\theta_{K_1}+M^{LL;K_{1B}}_{e\pi}\cos\theta_{K_1})\left[\xi_uC_1-\xi_t(C_3+C_9)\right]\non &&
-\xi_t(M^{LR;K_{1A}}_{e\pi}\sin\theta_{K_1}+M^{LR;K_{1B}}_{e\pi}\cos\theta_{K_1})(C_5+C_7)\non &&
+f_B(F^{LL;K_{1A}}_{a\pi}\sin\theta_{K_1}+F^{LL;K_{1B}}_{a\pi}\cos\theta_{K_1})\left[\xi_ua_2-\xi_t(a_4+a_{10})\right]\non &&
-\xi_tf_B(F^{SP;K_{1A}}_{a\pi}\sin\theta_{K_1}+F^{SP;K_{1B}}_{a\pi}\cos\theta_{K_1})(a_6+a_8). \label{kpi4}
\en
It is easy to get the total amplitudes for the decay modes including $\bar K_1(1400)^0$/$K_1(1400)^-$ by making the replacements with
$\sin\theta_{K_1}\to\cos\theta_{K_1}, \cos\theta_{K_1}\to -\sin\theta_{K_1}$ in Eqs.(\ref{kpi1}-\ref{kpi4}), respectively.
The total amplitudes for each $B\to K_1(1270)K, K_1(1400)K$ decay are given in Appendix A.
\section{Numerical results and discussions } \label{numer}

The input parameters in the numerical calculations \cite{pdg12,ckmfit} are listed as follows:
\be
f_B&=&210 MeV, f_{K_{1A}}=250 MeV, f^{\perp}_{K_{1B}}=190 MeV \\
\tau_{B^\pm}&=&1.638\times 10^{-12} s,\tau_{B^0}=1.525\times 10^{-12} s,\\
|V_{ud}|&=&0.974,  |V_{td}|=8.67\times10^{-3}, |V_{ub}|=3.51\times10^{-3}, \\
|V_{ts}|&=&0.0404, |V_{us}|=0.22534, , |V_{tb}|=0.999.
\en
Using the input parameters and the wave functions as specified in this section and Sec.\ref{proper}, it is easy to get the branching ratios for
the considered decays which are listed in Table \ref{bran1}, where the first error comes from the uncertainty in the $B$ meson shape
parameter $\omega_b=0.40\pm0.04$ GeV, the second error is from the
hard scale $t$, which we vary from $0.8t$ to $1.2t$, and the third error is from the combined uncertainties of the Gegenbauer moments
$a^{\perp}_1(K_{1A})=-1.08\pm0.48$ and $a^{\parallel}_1(K_{1B})=-1.95\pm0.45$.
\begin{table}
\caption{Branching ratios (in units of $10^{-6}$) for the decays
$B\to K_1(1270) \pi, K_1(1400)\pi$ and $B\to K_1(1270)K, K_1(1400)K$
for mixing angle $\theta_{\bar K_1}=-33^\circ$. Other model predictions
are also presented here for comparison. It is noticed that the
results of \cite{cmv} and \cite{vnp} are obtained for
mixing angle $32^{\circ}$, while those in \cite{cy} are obtained for mixing angle
$-37^{\circ}$.}
\begin{center}
\begin{tabular}{ccccccc|c}
\hline\hline   & \cite{cmv}  & \cite{vnp} &\cite{cy} & this work\\
\hline
$\bar B^0\to  K^{-}_1(1270)\pi^+$&$4.3$&$7.6$&$3.0^{+0.8+1.5+4.2}_{-0.6-0.9-1.4}$&$4.6^{+0.3+0.9+1.5}_{-0.1-0.8-1.2}$\\
$\bar B^0\to  \bar K^0_1(1270)\pi^0$ &$2.3$&$0.4$&$1.0^{+0.0+0.6+1.7}_{-0.0-0.3-0.6}$&$1.4^{+0.1+0.7+0.6}_{-0.1-0.5-0.5}$\\
$B^-\to  \bar K^0_1(1270)\pi^-$ &$4.7$&$5.8$&$3.5^{+0.1+1.8+5.1}_{-0.1-1.1-1.9}$&$3.5^{+0.4+1.9+1.6}_{-0.2-1.1-1.2}$\\
$B^-\to  K^{-}_1(1270)\pi^0$&$2.5$&$4.9$&$2.7^{+0.1+1.1+3.1}_{-0.1-0.7-1.0}$&$3.9^{+0.9+1.0+1.1}_{-0.5-0.7-1.0}$\\
$\bar B^0\to  K^{-}_1(1400)\pi^+$ &$2.3$&$4.0$&$5.4^{+1.1+1.7+9.9}_{-1.0-1.3-2.8}$&$3.0^{+0.5+0.1+0.9}_{-0.3-0.1-0.7}$\\
$\bar B^0\to  K^{0}_1(1400)\pi^0$ &$1.7$&$3.0$&$2.9^{+0.3+0.7+5.5}_{-0.3-0.6-1.7}$&$3.3^{+0.9+0.1+1.0}_{-0.7-0.0-0.8}$\\
$B^-\to  \bar K^0_1(1400)\pi^-$ &$2.5$&$3.0$&$6.5^{+1.0+2.0+11.6}_{-0.9-1.6-3.6}$&$5.0^{+1.3+1.0+1.4}_{-0.7-0.8-1.1}$\\
$B^-\to  K^{-}_1(1400)\pi^0$&$0.7$&$1.0$&$3.0^{+0.4+1.1+5.2}_{-0.4-0.7-1.3}$&$1.8^{+0.3+0.1+0.4}_{-0.2-0.2-0.3}$\\
\hline
$\bar B^0\to  K^{-}_1(1270)K^+$ &&&$0.01^{+0.01+0.00+0.02}_{-0.00-0.00-0.01}$&$0.13^{+0.01+0.00+0.23}_{-0.01-0.01-0.08}$\\
$\bar B^0\to  K^{+}_1(1270)K^-$ &&&$0.06^{+0.01+0.00+0.46}_{-0.01-0.00-0.06}$&$0.26^{+0.02+0.05+0.19}_{-0.02-0.04-0.12}$\\
$B^-\to  K^{0}_1(1270)K^-$ &$0.22$&&$0.25^{+0.01+0.15+0.39}_{-0.01-0.08-0.09}$&$1.11^{+0.01+0.19+0.43}_{-0.01-0.03-0.35}$\\
$B^-\to  K^{-}_1(1270)K^0$ & $0.02$&&$0.05^{+0.02+0.07+0.10}_{-0.02-0.03-0.04}$&$1.84^{+0.37+0.29+0.65}_{-0.28-0.25-0.42}$\\
$\bar B^0\to  \bar K^{0}_1(1270)K^0$ &$0.02$&&$2.30^{+0.16+1.13+1.43}_{-0.15-0.61-0.61}$&$1.71^{+0.34+0.27+0.51}_{-0.26-0.23-0.43}$\\
$\bar B^0\to  K^{0}_1(1270)\bar K^0$ &$0.20$&&$0.24^{+0.01+0.11+0.33}_{-0.01-0.07-0.13}$&$0.26^{+0.03+0.17+0.14}_{-0.06-0.01-0.08}$\\
$\bar B^0\to  K^{-}_1(1400)K^+$ &&&$0.09^{+0.01+0.00+0.23}_{-0.01-0.00-0.09}$&$0.64^{+0.14+0.00+0.13}_{-0.06-0.01-0.08}$\\
$\bar B^0\to  K^{+}_1(1400)K^-$ &&&$0.02^{+0.00+0.00+0.04}_{-0.00-0.00-0.00}$&$0.31^{+0.02+0.11+0.12}_{-0.00-0.01-0.09}$\\
$B^-\to  K^{0}_1(1400)K^-$ &$0.12$&&$0.48^{+0.08+0.15+0.81}_{-0.08-0.12-0.26}$&$0.90^{+0.13+0.11+1.21}_{-0.08-0.09-0.16}$\\
$B^-\to  K^{-}_1(1400)K^0$      &$4.4$&&$0.01^{+0.00+0.01+0.14}_{-0.00-0.00-0.01}$&$1.33^{+0.14+0.31+0.33}_{-0.10-0.22-0.22}$\\
$\bar B^0\to  \bar K^{0}_1(1400)K^0$ &$4.1$&&$0.08^{+0.01+0.17+0.59}_{-0.01-0.06-0.08}$&$1.46^{+0.16+0.31+0.33}_{-0.13-0.25-0.28}$\\
$\bar B^0\to  K^{0}_1(1400)\bar K^0$ &$0.11$&&$0.50^{+0.08+0.13+0.92}_{-0.07-0.11-0.32}$&$0.14^{+0.04+0.04+0.07}_{-0.03-0.03-0.02}$\\
\hline\hline
\end{tabular}\label{bran1}
\end{center}
\end{table}
From Table \ref{bran1} we can find that the branching ratios of $B\to K_1(1270)\pi, K_1(1400)\pi$ decays fall in $10^{-6}$ order.
The experimental data for the branching ratios of the decays $\bar B^0\to K_1(1270)^-\pi^+, K_1(1400)^-\pi^+$, which are given as
$(12.0\pm3.1^{+9.3}_{-4.5})\times10^{-6}$ and $(16.7\pm2.6^{+3.5}_{-5.0})\times10^{-6}$, respectively, are large and incompatible with all the present
theory predictions. Even for the two sided intervals $Br(\bar B^0\to K_1(1270)^-\pi^+)\in [0.6,2.5]\times10^{-5}$ and
$Br(\bar B^0\to K_1(1270)^-\pi^+)\in [0.8,2.4]\times10^{-5}$, they almost can not contain the different theoretical results. While the branching ratios of the
charged $B$ decays can be explained by the theories for the large uncertainties of the intervals $Br(B^-\to \bar K_1(1270)^0\pi^-)\in[0.0, 2.1]\times10^{-5},
Br(B^-\to \bar K_1(1400)^0\pi^-)\in[0.0, 2.5]\times10^{-5}$. The large large differences between theories and experiments do not happen to the decays
$\bar B^0\to a_1(1260)^\pm\pi^\mp$, which are tree-dominated. If the decay
constants $f_{a_1}, f_{\pi}$ and the form factors $V^{B\to a_1}_0, F^{B\to \pi}_0$ can be well determined, it is not difficult for us to
predict the branching ratios
of the decays $\bar B^0\to a_1(1260)^{\pm}\pi^{\mp}$ accurately, because the penguin contributions can be neglected and there are fewer uncertainties. For
the considered decays $\bar B^0\to K_1^\pm\pi^\mp$,
the tree operators are suppressed by the CKM matrix elements $V_{ub}V^*_{us}/(V_{cb}V^*_{cs})\sim 0.02$, and the penguin operators will play a significant role.
If the future data are really larger than the present predictions for here considered decays, the authors \cite{cy} claimed that there
are two possible reasons: one
is because the larger corrections from the weak annihilation and the hard spectator contributions, the other is from the charming penguin contributions.
In our calculations, the hard spectator contributions which correspond to the non-factorization emission diagram ones are very small. Although the
factorizable
annihilation contributions are more important, they can not promote the branching ratios too much. So we consider that the charming penguins are
more likely to explain the large data. Unfortunately, the charming penguins are non-perturbative in nature and remain untouched by
many theory approaches.
While it is helpful to consider these decays by using the
soft-collinear-effective-theory (SECT) \cite{bauer}, where the charming penguin contributions from loop diagrams are included. Certainly, these contributions
can also be incorporated in the final-state interactions \cite{hycheng1}. There exits the similar situation for the decays $\bar B^0\to a_1(1260)^+K^-, b_1(1235)^+K^-$
\cite{wwang}, where the PQCD predictions are larger than the data. The nonperturbative contributions, such as the final state interactions or the charming penguins,
are suggested to explain the data.
The penguin contributions from the factorization annihilation diagrams in the $K_{1B} \pi$ modes are much larger than those in the $K_{1A} \pi$ modes.
So we can find that the branching ratios of $B\to K_{1B} \pi$ decays are always larger than those of $B\to K_{1A} \pi$ decays, which is shown in Table \ref{branab}.
\begin{table}
\caption{Branching ratios (in units of $10^{-6}$) for the decays $B\to K_{1A} \pi, K_{1B}\pi$ and $B\to K_{1A}K, K_{1B}K$. The errors for these entries correspond
to the uncertainties from $\omega_B=0.4\pm0.04 GeV$, the hard scale $t$ varying from $0.8t$ to $1.2t$, and the Gegenbauer moments $a_1^{\perp}(K_{1A})=-1.08\pm0.48$ for $K_{1A}$ meson,
$a_1^{\parallel}(K_{1B})=-1.95\pm0.45$ for $K_{1B}$ meson, respectively.}
\begin{center}
\begin{tabular}{c|c|c|c}
\hline\hline
$\bar B^0\to  K^{-}_{1A}\pi^+$&$2.1^{+1.0+0.1+0.0}_{-0.6-0.1-0.3}$&$\bar B^0\to  K^{-}_{1B}\pi^+$&$5.6^{+0.1+0.8+2.1}_{-0.2-0.9-1.9}$\\
$\bar B^0\to  \bar K^0_{1A}\pi^0$ &$1.3^{+0.7+0.2+0.9}_{-0.5-0.2-0.6}$&$\bar B^0\to  \bar K^0_{1B}\pi^0$ &$3.4^{+0.1+1.0+1.1}_{-0.1-0.7-0.9}$\\
$B^-\to  \bar K^0_{1A}\pi^-$ &$3.9^{+1.9+0.6+1.7}_{-1.3-0.5-1.5}$&$B^-\to  \bar K^0_{1B}\pi^-$ &$4.7^{+0.2+2.2+1.8}_{-0.3-1.5-1.6}$\\
$B^-\to  K^{-}_{1A}\pi^0$&$2.1^{+0.9+0.2+0.6}_{-0.7-0.2-0.8}$&$B^-\to  K^{-}_{1B}\pi^0$&$3.7^{+0.1+0.7+1.2}_{-0.2-0.8-1.1}$\\
\hline
$\bar B^0\to  K^{-}_{1A}K^+$ &$0.47^{+0.03+0.00+0.28}_{-0.04-0.00-0.04}$&$\bar B^0\to  K^{-}_{1B}K^+$ &$0.34^{+0.04+0.01+0.14}_{-0.03-0.01-0.07}$\\
$\bar B^0\to  K^{+}_{1A}K^-$ &$0.14^{+0.01+0.01+0.11}_{-0.00-0.01-0.13}$&$\bar B^0\to  K^{+}_{1B}K^-$ &$0.38^{+0.03+0.03+0.26}_{-0.03-0.02-0.19}$\\
$B^-\to  K^{0}_{1A}K^-$ &$1.24^{+0.13+0.08+1.74}_{-0.12-0.07-0.65}$&$B^-\to  K^{0}_{1B}K^-$ &$0.60^{+0.04+0.19+0.13}_{-0.04-0.12-0.08}$\\
$B^-\to  K^{-}_{1A}K^0$ &$0.29^{+0.02+0.05+1.26}_{-0.01-0.03-0.03}$&$B^-\to  K^{-}_{1B}K^0$ &$2.65^{+0.53+0.48+0.67}_{-0.34-0.41-0.57}$\\
$\bar B^0\to  \bar K^{0}_{1A}K^0$ &$0.10^{+0.00+0.05+0.10}_{-0.00-0.03-0.04}$&$\bar B^0\to  \bar K^{0}_{1B}K^0$ &$2.71^{+0.30+0.52+0.66}_{-0.30-0.43-0.58}$\\
$\bar B^0\to  K^{0}_{1A}\bar K^0$ &$0.16^{+0.12+0.06+0.18}_{-0.06-0.03-0.10}$&$\bar B^0\to  K^{0}_{1B}\bar K^0$ &$0.17^{+0.01+0.08+0.09}_{-0.01-0.05-0.06}$\\
\hline\hline
\end{tabular}\label{branab}
\end{center}
\end{table}

For the decays $B\to K_1(1270)K, K_1(1400)K$, there are no experimental data or upper limits up to now.
Although the decays $\bar B^0\to K_1^{\pm}K^{\mp}$
can occur only via annihilation type diagrams, their branching ratios might not be so small as those predicted by the QCDF approach. If our predictions can be
confirmed by the future LHCb or the super B experiments, one can say that the PQCD approach is one of the few methods, which can
be used to quantitatively calculate the
annihilation type contributions. In the previous years both the experimenters and the theorists considered that
the branching ratio of $B^0\to K^+K^-$ was at $10^{-8}$ order, but two years ago the CDF and LHCb collaborations gave their first measurements of this decay by
$(2.3\pm1.0\pm1.0)\times10^{-7}$ \cite{cdf} and $(1.3^{+0.6}_{-0.5}\pm0.7)\times 10^{-7}$ \cite{lhcb}, respectively. Later, these results are confirmed by the PQCD recalculated
result $1.56\times10^{-7}$ \cite{xiao} without introducing too much uncertainties. It shows that the PQCD approach can determine correctly the strength of
penguin-annihilation amplitudes. Whether the PQCD approach can give  reasonable predictions for the pure annihilation decays
$\bar B^0\to K_1(1270)^{\pm}K^{\mp},K_1(1400)^{\pm}K^{\mp}$ also deserves our attention and research.
For the decay $\bar B^0\to K^0_{1B}\bar K^0$ can not receive a large emission factorization amplitude, because of the small decay constant $f_{K_{1B}}$ compared
with $f_{K_{1A}}$, while it has a large annihilation factorization amplitude, which makes its branching ratio slightly larger than that of
$\bar B^0\to K^0_{1A}\bar K^0$. The branching ratios of these two decays are at the order of $10^{-7}$. But it is very different to the decay
$\bar B^0\to \bar K^0_{1B}K^0$: Except having a large annihilation factorization amplitude, it can also obtain a large emission factorization amplitude
at the same time, because here the emission meson is $K^0$ with a larger decay constant $f_{K}=0.16$. So this decay gets a large branching ratio,
which amounts to $2.71\times10^{-6}$.
Even though the decay $\bar B^0\to \bar K^0_{1A}K^0$ has a small branching ratio, the physical final states $\bar K_1(1200)^0K^0, \bar K_1(1400)^0K^0$, which are
mixes of the former two group flavor states, still might get a large branching ratio. It has been verified by the different theories, which are shown in Table
\ref{bran1}. But the branching ratio of the decay
$\bar B^0\to\bar K_1(1400)^0K^0$ predicted by the QCDF approach seems too small compared with the results given by the PQCD and the naive factorization approaches, which can be
clarified by the future experiments. There exists the similar situation for the decay $B^-\to K_1(1400)^-K^0$. Another decay channel, where exists large
divergence between the predictions, is $B^-\to K_1(1200)^-K^0$. The Feynman diagrams of this decay can be obtained from those of the decay $\bar B^0\to\bar K_1(1200)^0K^0$ by
replacing the spectator quark $d$ with $u$, so the difference of the branching ratios of these two decays should not be so large. In a word,
the branching ratios of the charged $B$ decays are at or near the order of $10^{-6}$, those of the pure annihilation decays are at
the order of $10^{-7}$ by taking the mixing angle $\theta_{K_1}=33^\circ$.

In order to compare with other theoretical predictions, we also list the
branching ratios with the mixing angle $\theta_{\bar
K_1}=-58^\circ$ shown in Table \ref{bran2}. One can find that the
branching ratios of the decays $B^-\to K_1^-(1270)K^0, \bar B^0\to \bar K_1^0(1270)K^0$ have a remarkable decrease from the mixing angles $-33^\circ$ to
$-58^\circ$, while those of the decays $B^-\to K_1^-(1400)K^0, \bar B^0\to \bar K_1^0(1400)K^0$ have a remarkable increase.

\begin{table}
\caption{Same as Table\ref{bran1} except for the mixing angle
$\theta_{\bar K_1}=-58^\circ$.}
\begin{center}
\begin{tabular}{ccccccc|c}
\hline\hline   & \cite{cmv}  & \cite{vnp} &\cite{cy} & this work\\
\hline
$\bar B^0\to  K^{-}_1(1270)\pi^+$&$4.3$&$7.6$&$2.7^{+0.6+1.3+4.4}_{-0.5-0.8-1.5}$&$3.2^{+0.7+0.5+0.8}_{-0.5-0.5-0.8}$\\
$\bar B^0\to  \bar K^0_1(1270)\pi^0$ &$2.1$&$0.4$&$0.8^{+0.1+0.5+1.7}_{-0.1-0.3-0.6}$&$0.5^{+0.2+0.0+0.4}_{-0.0-0.2-0.2}$\\
$B^-\to  \bar K^0_1(1270)\pi^-$ &$4.7$&$5.8$&$3.0^{+0.2+0.1+2.7}_{-0.2-0.2-2.2}$&$3.2^{+1.3+1.2+1.3}_{-0.9-0.8-1.2}$\\
$B^-\to  K^{-}_1(1270)\pi^0$&$1.6$&$4.9$&$2.5^{+0.1+1.0+3.2}_{-0.1-0.7-1.0}$&$3.3^{+1.1+0.7+0.8}_{-0.8-0.6-1.1}$\\
$\bar B^0\to  K^{-}_1(1400)\pi^+$ &$2.3$&$4.0$&$2.2^{+1.1+0.7+2.6}_{-0.8-0.6-1.3}$&$4.5^{+0.0+0.3+1.5}_{-0.0-0.5-1.3}$\\
$\bar B^0\to  K^{0}_1(1400)\pi^0$ &$1.6$&$1.7$&$1.5^{+0.4+0.3+1.7}_{-0.3-0.3-0.9}$&$4.1^{+0.8+0.7+1.2}_{-0.4-0.4-0.8}$\\
$B^-\to  \bar K^0_1(1400)\pi^-$ &$2.5$&$3.0$&$2.8^{+1.0+0.9+3.0}_{-0.8-0.9-1.7}$&$5.4^{+0.3+1.6+1.5}_{-0.2-1.2-1.4}$\\
$B^-\to  K^{-}_1(1400)\pi^0$&$0.6$&$1.4$&$1.0^{+0.4+0.4+1.2}_{-0.3-0.4-0.5}$&$2.5^{+0.0+0.3+0.8}_{-0.0-0.4-0.7}$\\
\hline
$\bar B^0\to  K^{-}_1(1270)K^+$ &&&$0.01^{+0.00+0.00+0.02}_{-0.00-0.00-0.01}$&$0.19^{+0.01+0.00+0.37}_{-0.01-0.00-0.09}$\\
$\bar B^0\to  K^{+}_1(1270)K^-$ &&&$0.04^{+0.01+0.00+0.27}_{-0.01-0.00-0.04}$&$0.16^{+0.00+0.02+0.12}_{-0.02-0.03-0.06}$\\
$B^-\to  K^{0}_1(1270)K^-$ &$0.22$&&$0.22^{+0.01+0.12+0.39}_{-0.01-0.07-0.12}$&$1.47^{+0.10+0.16+1.59}_{-0.06-0.10-0.58}$\\
$B^-\to  K^{-}_1(1270)K^0$ & $0.75$&&$0.05^{+0.02+0.09+0.10}_{-0.01-0.03-0.04}$&$0.78^{+0.17+0.09+0.97}_{-0.13-0.08-0.19}$\\
$\bar B^0\to  \bar K^{0}_1(1270)K^0$ &$0.70$&&$2.10^{+0.13+1.23+1.31}_{-0.13-0.65-0.57}$&$0.46^{+0.13+0.07+0.17}_{-0.09-0.05-0.13}$\\
$\bar B^0\to  K^{0}_1(1270)\bar K^0$ &$0.20$&&$0.26^{+0.10+0.12+0.47}_{-0.01-0.08-0.17}$&$0.23^{+0.09+0.13+0.18}_{-0.06-0.08-0.16}$\\
$\bar B^0\to  K^{-}_1(1400)K^+$ &&&$0.07^{+0.02+0.00+0.16}_{-0.02-0.00-0.06}$&$0.58^{+0.06+0.01+0.15}_{-0.06-0.01-0.13}$\\
$\bar B^0\to  K^{+}_1(1400)K^-$ &&&$0.01^{+0.00+0.00+0.16}_{-0.02-0.00-0.06}$&$0.42^{+0.03+0.01+0.22}_{-0.02-0.00-0.16}$\\
$B^-\to  K^{0}_1(1400)K^-$ &$0.12$&&$0.22^{+0.07+0.07+0.24}_{-0.07-0.07-0.13}$&$0.54^{+0.04+0.14+0.76}_{-0.02-0.11-0.13}$\\
$B^-\to  K^{-}_1(1400)K^0$      &$3.9$&&$0.01^{+0+0.02+0.04}_{-0-0.00-0.00}$&$2.39^{+0.34+0.50+0.48}_{-0.25-0.39-0.48}$\\
$\bar B^0\to  \bar K^{0}_1(1400)K^0$ &$3.6$&&$0.10^{+0.02+0.21+0.15}_{-0.02-0.08-0.10}$&$2.24^{+0.36+0.40+0.59}_{-0.28-0.34-0.51}$\\
$\bar B^0\to  K^{0}_1(1400)\bar K^0$ &$0.11$&&$0.25^{+0.07+0.08+0.31}_{-0.07-0.07-0.15}$&$0.21^{+0.02+0.13+0.09}_{-0.01-0.07-0.07}$\\
\hline\hline
\end{tabular}\label{bran2}
\end{center}
\end{table}

\begin{figure}[t,b]
\begin{center}
\includegraphics[scale=0.7]{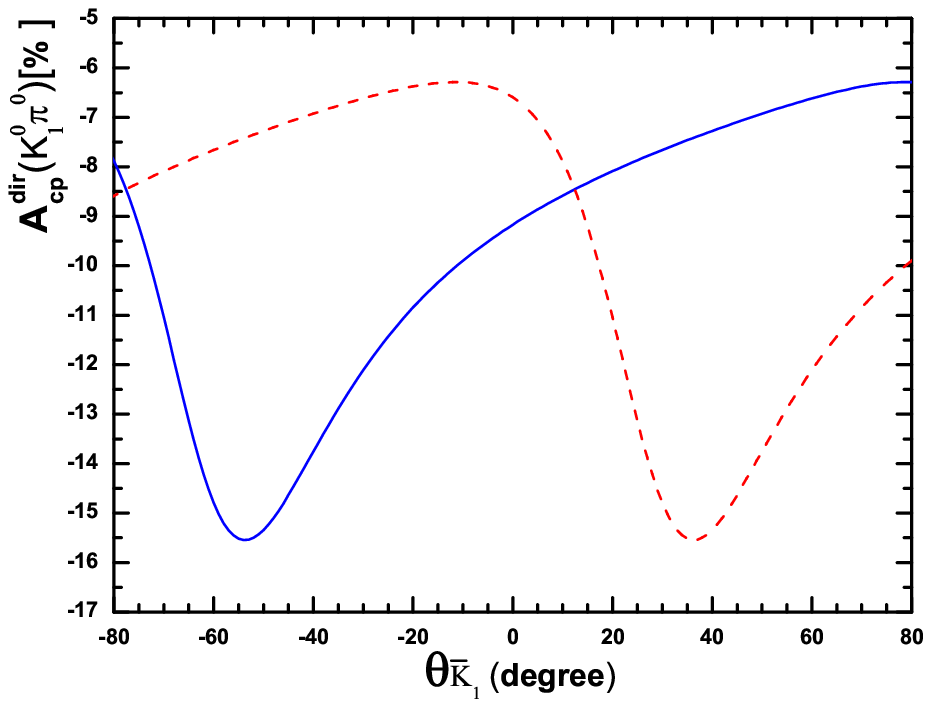}
\includegraphics[scale=0.7]{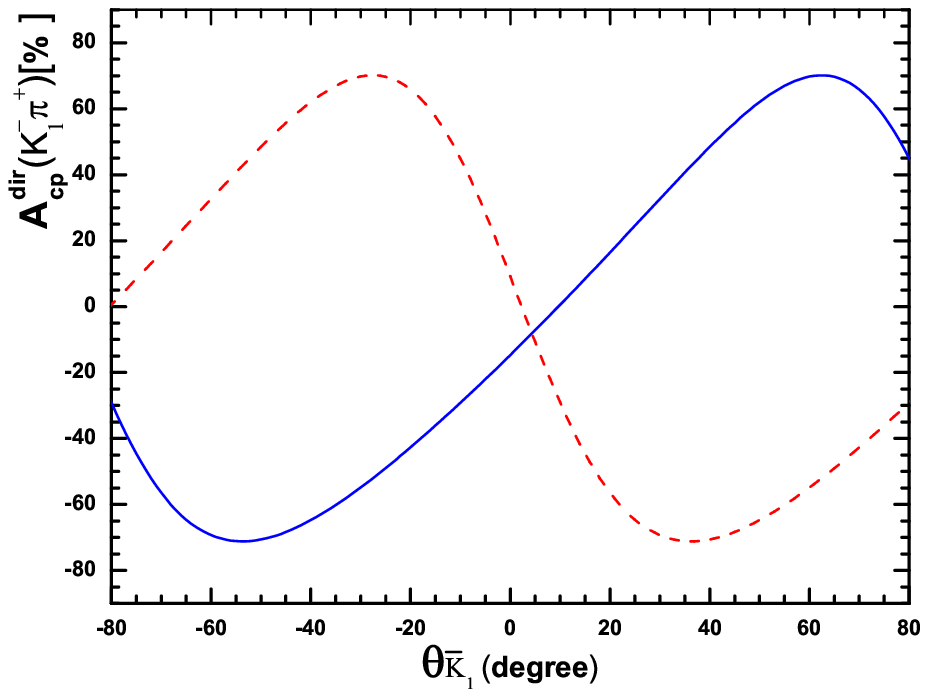}
\vspace{0.3cm} \caption{The dependence of the direct CP-violating
asymmetries on the mixing angle $\theta_{\bar K_1}$: the solid lines represent
the decays $\bar B^0\to  K_1(1270)^{0}\pi^0$ (left), $\bar B^0\to  K_1(1270)^{-}\pi^+$ (right), and the dashed lines
are for the decays $\bar B^0\to  K_1(1400)^{0}\pi^0$ (left), $\bar B^0\to  K_1(1400)^{-}\pi^+$ (right),
respectively.}\label{fig2}
\end{center}
\end{figure}
\begin{figure}[t,b]
\begin{center}
\includegraphics[scale=0.7]{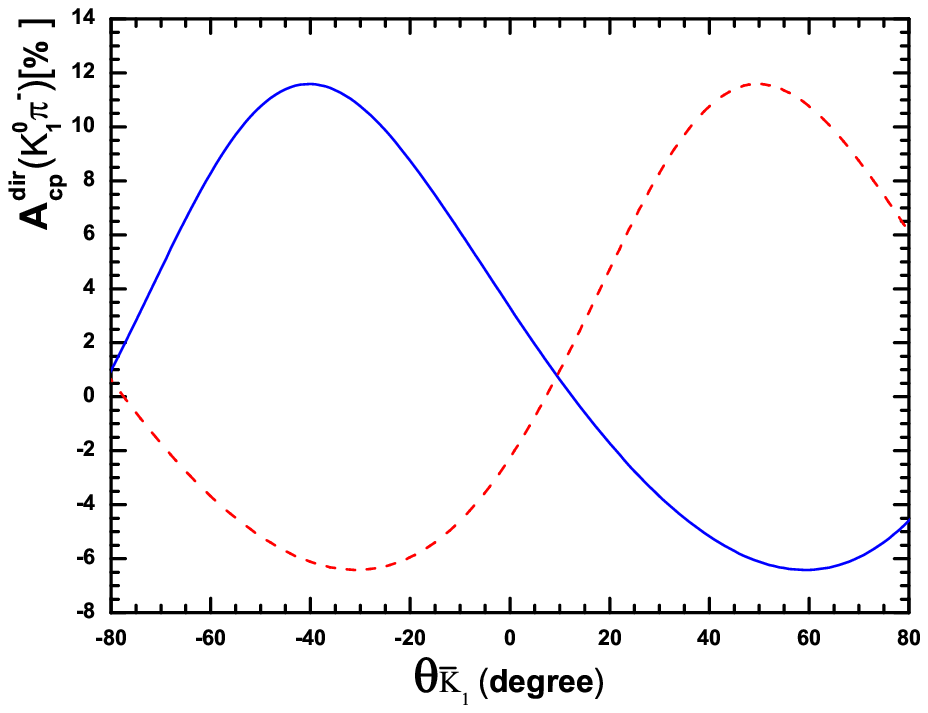}
\includegraphics[scale=0.7]{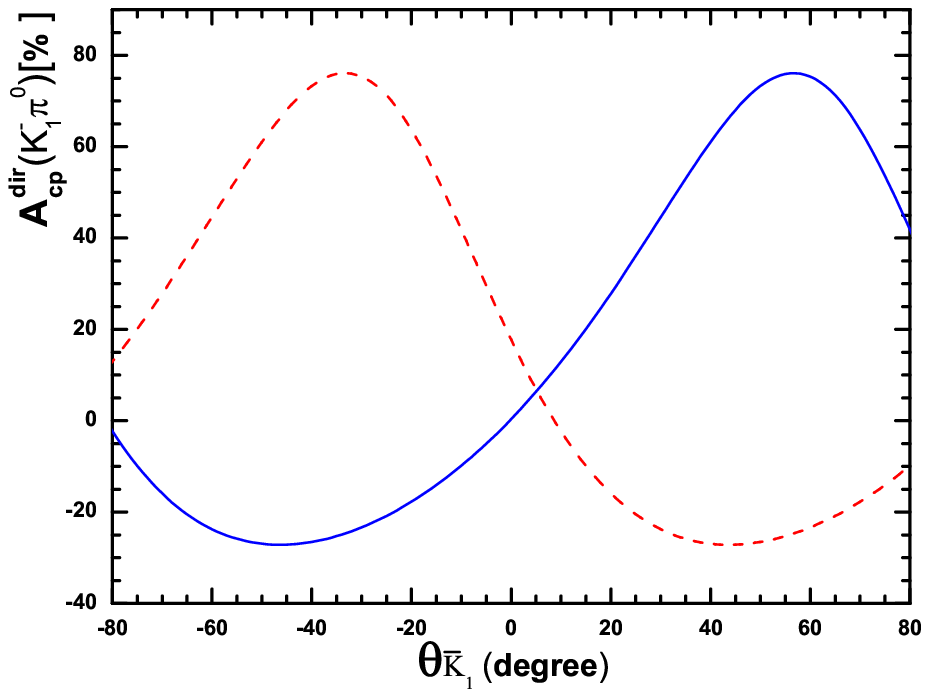}
\vspace{0.3cm} \caption{The dependence of the direct CP-violating
asymmetries on the mixing angle $\theta_{\bar K_1}$: the solid lines represent
the decays $B^-\to  K_1(1270)^{0}\pi^-$ (left), $B^-\to  K_1(1270)^{-}\pi^0$ (right), and the dashed lines
are for the decays $B^-\to  K_1(1400)^{0}\pi^-$ (left), $B^-\to  K_1(1400)^{-}\pi^0$ (right),
respectively.}\label{fig3}
\end{center}
\end{figure}
\begin{figure}[t,b]
\begin{center}
\includegraphics[scale=0.7]{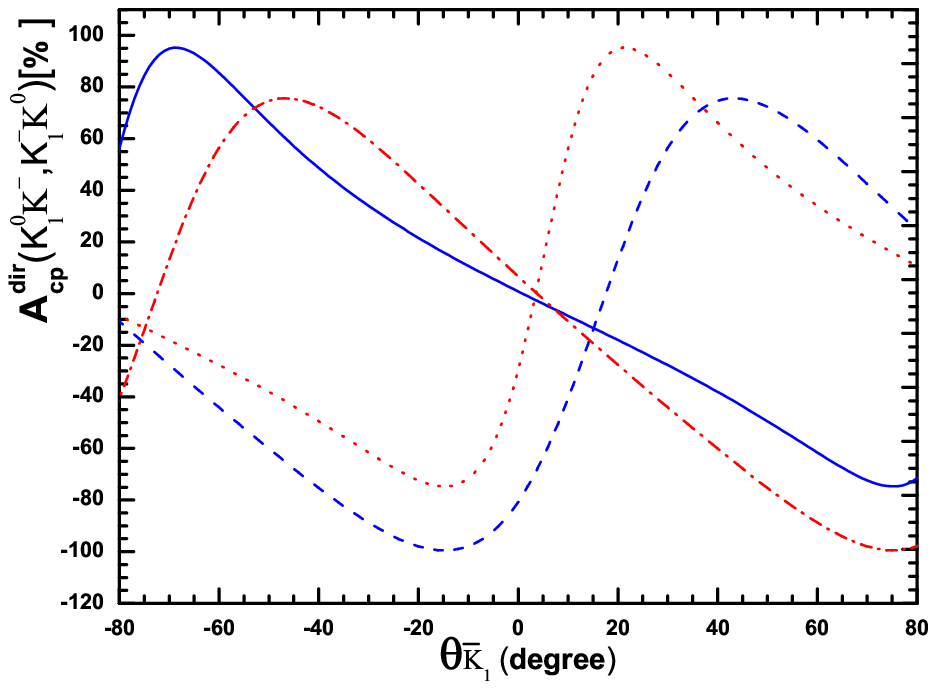}
\includegraphics[scale=0.7]{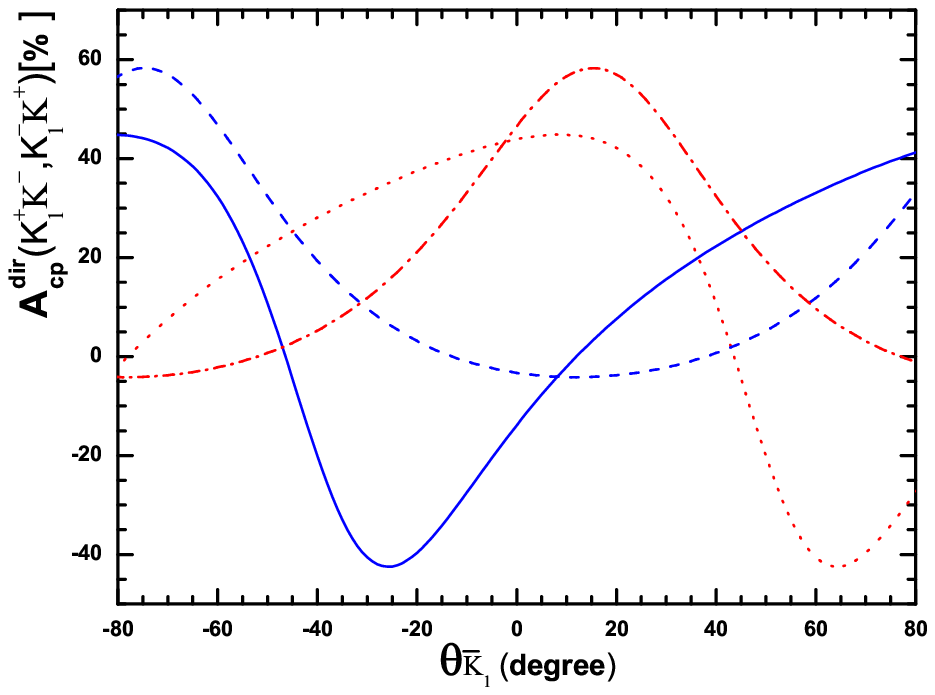}
\vspace{0.3cm} \caption{The dependence of the direct CP-violating
asymmetries on the mixing angle $\theta_{\bar K_1}$: the solid lines represent
the decays $B^-\to  K_1(1270)^{-}K^0$ (left), $\bar B^0\to  K_1(1270)^{-}K^+$ (right), the dashed lines
are for the decays $B^-\to  K_1(1270)^{0}K^-$ (left), $\bar B^0 \to  K_1(1270)^{+}K^-$ (right), the dot lines
are for the decays $B^-\to  K_1(1400)^{-}K^0$ (left), $B^-\to  K_1(1400)^{-}K^+$ (right), and the dash-dot lines represent the decays
$B^-\to  K_1(1400)^{0}K^-$ (left), $\bar B^0\to  K_1(1400)^{+}K^-$ (right),
respectively.}\label{fig4}
\end{center}
\end{figure}
\begin{table}
\caption{Direct CP violation (in units of $\%$) for the decays $B\to
K_{1A} \pi, K_{1B}\pi$ and $B\to K_{1A}K, K_{1B}K$. The errors for these entries correspond
to the uncertainties from $\omega_B=0.4\pm0.04$ GeV, the hard scale $t$ varying from $0.8t$ to $1.2t$, and the Gegenbauer moment $a_1^{\perp}(K_{1A})=-1.08\pm0.48$ for $K_{1A}$ meson,
$a_1^{\parallel}(K_{1B})=-1.95\pm0.45$ for $K_{1B}$ meson, respectively..}
\begin{center}
\begin{tabular}{c|c|c|c}
\hline\hline
$\bar B^0\to  K^{-}_{1A}\pi^+$&$9.1^{+2.4+0.8+3.0}_{-2.0-0.8-3.4}$&$\bar B^0\to  K^{-}_{1B}\pi^+$&$-14.7^{+1.2+0.0+1.1}_{-1.4-0.2-1.6}$\\
$\bar B^0\to  \bar K^0_{1A}\pi^0$ &$-6.6^{+1.3+0.9+2.8}_{-1.4-1.0-8.4}$&$\bar B^0\to  \bar K^0_{1B}\pi^0$ &$-9.2^{+1.0+3.3+1.6}_{-0.7-3.5-1.9}$\\
$B^-\to  \bar K^0_{1A}\pi^-$ &$-2.3^{+0.8+0.8+1.5}_{-1.2-0.6-6.8}$&$B^-\to  \bar K^0_{1B}\pi^-$ &$3.3^{+0.1+0.6+1.9}_{-0.1-0.6-1.3}$\\
$B^-\to  K^{-}_{1A}\pi^0$&$17.7^{+4.1+3.0+17.1}_{-3.5-3.1-7.4}$&$B^-\to  K^{-}_{1B}\pi^0$&$3.4^{+1.2+0.0+0.0}_{-1.4-4.6-6.8}$\\
\hline
$\bar B^0\to  K^{-}_{1A}K^+$ &$43.9^{+1.7+0.5+0.0}_{-1.3-3.1-35.6}$&$\bar B^0\to  K^{-}_{1B}K^+$ &$-13.9^{+2.5+1.8+0.4}_{-2.6-2.0-0.4}$\\
$\bar B^0\to  K^{+}_{1A}K^-$ &$46.5^{+0.5+4.4+40.3}_{-1.3-3.3-29.5}$&$\bar B^0\to  K^{+}_{1B}K^-$ &$-3.3^{+1.1+6.8+1.6}_{-0.7-4.1-1.7}$\\
$B^-\to  K^{0}_{1A}K^-$ &$6.6^{+1.6+3.1+4.9}_{-1.7-3.8-1.8}$&$B^-\to  K^{0}_{1B}K^-$ &$-80.7^{+1.3+4.4+11.1}_{-1.7-3.5-2.9}$\\
$B^-\to  K^{-}_{1A}K^0$ &$-29.4^{+7.6+2.6+86.7}_{-6.3-1.8-0.0}$&$B^-\to  K^{-}_{1B}K^0$ &$0.8^{+2.7+0.4+4.0}_{-3.6-0.5-2.9}$\\
\hline\hline
\end{tabular}\label{dircp}
\end{center}
\end{table}
Now we turn to the evaluations of the CP-violating asymmetries in the PQCD approach.
For the neutral $\bar B^0$ (the charged $B^{-}$) decays the direct CP-violating asymmetries can be defined as
\be
\acp^{dir}&=&\frac{  \Gamma(\bar B^0(B^-)\to f)-\Gamma(B^0(B^+)\to \bar f)}{
 \Gamma(\bar B^0(B^-)\to f)+\Gamma(B^0(B^+)\to \bar f)}=\frac{2z\sin\theta\sin\delta}
{(1+2z\cos\theta\cos\delta+z^2) }\;, \label{cpva}
\en
where $\delta$ is the relative strong phase between the tree and penguin amplitudes, and $\theta$ the CKM weak phase $\theta=\alpha$ for $b\to d$ transition,
$\theta=\gamma$ for $b\to s$ transition. Certainly, if the final states are the same for $B^0$ and $\bar B^0$, that is $f=\bar f$, the CP-asymmetries may be
time-dependent, including not only the direct CP violation but also the mixing-induced CP violation.
Using the input parameters and the wave functions as specified in this section and Sec.\ref{proper}, it is easy to get
the PQCD predictions (in units of $10^{-2}$) for the direct CP-violating asymmetries of $B$ decaying to each
flavor final state, which are listed in Table \ref{dircp}. For the real physical final states, which are mixes of the corresponding flavor states, their direct
CP-violating asymmetries will be dependent on the mixing angle $\theta_{\bar K_1}$. As has been emphasised before, $\theta_{\bar K_1}$ for the antiparticle states
$\bar K_{1}(1270), \bar K_{1}(1400)$ is of opposite sign to that for the particle states $K_{1}(1270), K_1(1400)$. For taking the convention of
decay constant $f_{K_{1B}}$ in this work, so $\theta_{K_1}$ is positive and $\theta_{\bar K_1}$ is negative. In Fig.\ref{fig2}-Fig.\ref{fig4}, we give the
dependence of the direct CP-violating asymmetries on the mixing angle $\theta_{\bar K_1}$ for each decay. Here taking $\theta_{\bar K_1}=-33^\circ$ or $\theta_{\bar K_1}=-58^\circ$,
we can read each direct CP-violating asymmetry from these figures. It is noticed that for the decays $\bar B^0\to K_1(1270)^+K^-, K_1(1400)^+K^-$,
$B^-\to K_1(1270)^0K^-, K_1(1400)^0K^-$, which include the particle states, their direct CP-violating asymmetry values are still read at $-33^\circ$ or
$-58^\circ$ for $\theta_{K_1}=-\theta_{\bar K_1}$ and so the corresponding mixing angle is positive.
The signs of the direct CP-violating asymmetries of
$B\to K_1(1270)K(\pi)$ and $B\to K_1(1400)K(\pi)$ are opposite at the mixing angle $\theta_{\bar K_1}=-33^\circ$ for most
of these decays except only two groups, whose
direct CP-violating asymmetries are predicted as $\acp^{dir}(\bar B^0\to \bar K_1(1270)^0\pi^0)=-12.6\%, \acp^{dir}(\bar B^0\to \bar K_1(1400)^0\pi^0)=-6.7\%$ and
$\acp^{dir}(\bar B^0\to K_1(1270)^+ K^-)=12.2\%, \acp^{dir}(\bar B^0\to K_1(1400)^+ K^-)=9.6\%$, respectively.
From Table \ref{dircp}, one can find that the direct CP-violating asymmetries of each decay $B\to K_{1A}\pi, K_{1B}\pi$ are not large, while those for some
real physical final states become very large. For example, the direct CP-violating asymmetries of the decays
$\bar B^0\to K_1(1270)^-\pi^+, K_1(1400)^-\pi^+$ are about $-58.1\%$ and $68.4\%$ at the mixing angle $-33^\circ$, respectively.
Certainly, we only learn phenomenally about the mixing angle $\theta_{K_1}$ at present and have no accurate calculations or measurements. Furthermore, the direct
CP-violating asymmetries are sensitive to the mixing angle. It is much
more complex for some considered decays where the nonperturbative contributions, such as charming penguins, give large corrections,
and the corresponding direct CP-violating asymmetries
may also change. So we can't confirm that these decays must have so large direct CP-violating asymmetries.
As for the decays $\bar B^0\to \bar K_1(1270)^0K^0, \bar K_1(1400)^0K^0$, there is no tree contribution at the leading order, so the direct CP-violating asymmetry
is naturally zero.
\section{Conclusion}\label{summary}

In this paper, by using the decay constants and the light-cone distribution amplitudes
derived from the QCD sum-rule method, we research the decays $B\to K_1(1270)\pi(K), K_1(1400)\pi(K)$
in the PQCD approach and find that
\begin{itemize}
\item
All the theoretical predictions for the branching ratios of the decays $\bar B^0\to K_1(1270)^+\pi^-, K_1(1400)^+\pi^-$ are
incompatible with the present experimental data. There exists the similar situation for the decays $\bar B^0\to a_1(1260)^+K^-, b_1(1235)^+K^-$, where
the nonperturbative contributions, such as the final state interactions or the charming penguins, are needed to explain the data. But the difference is that
the nonperturbative
contributions seem to play opposite roles in these two groups of decays. If the future data are really larger than the present predictions for some
considered decays, it might indicate that the nonperturbative contributions have pronounced corrections for some decay channels which include the higher
resonances in the final states.
\item
The pure annihilation type decays $\bar B^0\to K_1^{\pm}(1270)K^{\mp}, K_1^{\pm}(1400)K^{\mp}$ are good channels to test whether an approach can be used to
calculate correctly the strength of the
penguin-annihilation amplitudes. Their branching ratios are predicted at $10^{-7}$ order.
\item
In the four final neutral flavor states $K^0_{1A}\bar K^0, K^0_{1B}\bar K^0, \bar K^0_{1A} K^0, \bar K^0_{1B} K^0$, the decay $\bar B^0\to\bar K^0_{1B} K^0$ have
the largest branching ratio which is of $10^{-6}$ order, while the other decays with the branching ratios at $10^{-7}$ order. So the decays
$\bar B^0\to \bar K_1(1200)^0K^0, \bar K_1(1400)K^0$ which include the real physical states can have large branching ratios at the mixing angle $\theta_{\bar K_1}=-33^\circ$ compare with the decays
$\bar B^0\to  K_1(1200)^0\bar K^0, K_1(1400)\bar K^0$.
\item
The signs of the direct CP-violating asymmetries are opposite between almost of the decays $B\to K_1(1270)K(\pi)$ and $B\to K_1(1400)K(\pi)$ at mixing angle
$\theta_{K_1}=-33^\circ$ except only two groups, whose direct CP-violating asymmetries are predicted as $\acp^{dir}(\bar B^0\to \bar K_1(1270)^0\pi^0)=-12.6\%, \acp^{dir}(\bar B^0\to \bar K_1(1400)^0\pi^0)=-6.7\%$ and
$\acp^{dir}(\bar B^0\to K_1(1270)^+ K^-)=12.2\%, \acp^{dir}(\bar B^0\to K_1(1400)^+ K^-)=9.6\%$, respectively.
\item
The strong phase introduced by the nonperturbative contributions might produce dramatic effects on some of the considered decays, such as
$\bar B^0\to K_1(1270)^-\pi^+, K_1(1400)^-\pi^+,K_1(1270)^-\pi^0,K_1(1270)^-\pi^0$, and these effects could exceed those from the parametric
uncertainties in the case of the CP asymmetries.
\end{itemize}

\section*{Acknowledgment}
This work is partly supported by the National Natural Science Foundation of China under Grant No. 11147004, No.11147008, No.11347030, by the Program of the Youthful Key
Teachers in Universities of Henan Province under Grants No. 001166, and by the Program for Science and Technology Innovation Talents in Universities of Henan
Province 14HASTIT037. The authors would like to thank Prof. Hai-Yang Cheng and Prof. Cai-Dian Lu for helpful discussions.
\begin{appendix}
\section{Analytic formulas for the decay amplitudes }\label{sec:aa}
\be
A(K_1(1270)^{0}\bar K^0)&=&-\xi_t(f_{K_{1A}}\sin\theta_{K_1}+f_{K_{1B}}\cos\theta_{K_1})F^{LL}_{eK}(a_4-\frac{1}{2}a_{10})\non &&
-\xi_t(M^{LL;K_{1A}}_{eK}\sin\theta_{K_1}+M^{LL;K_{1B}}_{eK}\cos\theta_{K_1})(C_3-\frac{1}{2}C_9)\non &&
-\xi_t(M^{LR;K_{1A}}_{eK}\sin\theta_{K_1}+M^{LR;K_{1B}}_{eK}\cos\theta_{K_1})(C_5-\frac{1}{2}C_7)\non &&
-\xi_t(M^{LL;K_{1A}}_{aK}\sin\theta_{K_1}+M^{LL;K_{1B}}_{aK}\cos\theta_{K_1})(C_3-\frac{1}{2}C_9)\non &&
-\xi_t(M^{LL;K_{1A}}_{aK}\sin\theta_{K_1}+M^{LL;K_{1B}}_{aK}\cos\theta_{K_1})(C_4-\frac{1}{2}C_{10})\non &&
-\xi_t(M^{LR;K_{1A}}_{aK}\sin\theta_{K_1}+M^{LR;K_{1B}}_{aK}\cos\theta_{K_1})(C_5-\frac{1}{2}C_7)\non &&
-\xi_t(M^{SP;K_{1A}}_{aK}\sin\theta_{K_1}+M^{SP;K_{1B}}_{aK}\cos\theta_{K_1})(C_6-\frac{1}{2}C_8 )\non &&
-\xi_tf_B(F^{LL;K_{1A}}_{aK}\sin\theta_{K_1}+F^{LL;K_{1B}}_{aK}\cos\theta_{K_1})(a_3-\frac{1}{2}a_{9})\non &&
-\xi_tf_B(F^{LL;K_{1A}}_{aK}\sin\theta_{K_1}+F^{LL;K_{1B}}_{aK}\cos\theta_{K_1})(a_4-\frac{1}{2}a_{10})\non &&
-\xi_tf_B(F^{LL;K_{1A}}_{aK}\sin\theta_{K_1}+F^{LL;K_{1B}}_{aK}\cos\theta_{K_1})(a_5-\frac{1}{2}a_{7})\non &&
-\xi_tf_B(F^{SP;K_{1A}}_{aK}\sin\theta_{K_1}+F^{SP;K_{1B}}_{aK}\cos\theta_{K_1})(a_6-\frac{1}{2}a_8)\non &&
-\xi_t(M^{LL;K}_{aK_{1A}}\sin\theta_{K_1}+M^{LL;K}_{aK_{1B}}\cos\theta_{K_1})(C_4-\frac{1}{2}C_{10})\non &&
-\xi_t(M^{SP;K}_{aK_{1A}}\sin\theta_{K_1}+M^{SP;K}_{aK_{1B}}\cos\theta_{K_1})(C_6-\frac{1}{2}C_8 )\non &&
-\xi_tf_B(F^{LL;K}_{aK_{1A}}\sin\theta_{K_1}+F^{LL;K}_{aK_{1B}}\cos\theta_{K_1})(a_3-\frac{1}{2}a_{9})\non &&
-\xi_tf_B(F^{LL;K}_{aK_{1A}}\sin\theta_{K_1}+F^{LL;K}_{aK_{1B}}\cos\theta_{K_1})(a_5-\frac{1}{2}a_{7}), \label{kk1}
\en
\be
A(K_1(1270)^{0}K^-)&=&-\xi_t(f_{K_{1A}}\sin\theta_{K_1}+f_{K_{1B}}\cos\theta_{K_1})F^{LL}_{eK}(a_4-\frac{1}{2}a_{10})\non &&
-\xi_t(M^{LL;K_{1A}}_{eK}\sin\theta_{K_1}+M^{LL;K_{1B}}_{eK}\cos\theta_{K_1})(C_3-\frac{1}{2}C_9)\non &&
-\xi_t(M^{LR;K_{1A}}_{eK}\sin\theta_{K_1}+M^{LR;K_{1B}}_{eK}\cos\theta_{K_1})(C_5-\frac{1}{2}C_7)\non &&
+(M^{LL;K_{1A}}_{aK}\sin\theta_{K_1}+M^{LL;K_{1B}}_{aK}\cos\theta_{K_1})(\xi_u C_1-\xi_t(C_3+C_9))\non &&
-\xi_t(M^{LR;K_{1A}}_{aK}\sin\theta_{K_1}+M^{LR;K_{1B}}_{aK}\cos\theta_{K_1})(C_5+C_7)\non &&
+f_B(F^{LL;K_{1A}}_{aK}\sin\theta_{K_1}+F^{LL;K_{1B}}_{aK}\cos\theta_{K_1})(\xi_ua_2-\xi_t(a_4+a_{10})\non &&
-\xi_tf_B(F^{SP;K_{1A}}_{aK}\sin\theta_{K_1}+F^{SP;K_{1B}}_{aK}\cos\theta_{K_1})(a_6+a_8). \label{kk2}
\en
In the upper two formulae, if changing the first term as $-\xi_tf_{K}(F^{LL}_{eK_{1A}}\sin\theta_{K_1}+F^{LL}_{eK_{1B}}\cos\theta_{K_1})(a_4-\frac{1}{2}a_{10}))
-\xi_tf_K(F^{SP}_{eK_{1A}}\sin\theta_{K_1}+F^{SP}_{eK_{1B}}\cos\theta_{K_1})(a_6-\frac{1}{2}a_8)$, and at the same time exchanging the positions of
$K_{1A}(K_{1B})$ and $K$ in other terms, we will get the decay amplitudes of $\bar B^0\to\bar K_1(1270)^{0} K^0$ and $B^-\to K_1(1270)^{-} K^0$, respectively.
\be
A(K_1(1270)^{+} K^-)&=&
(M^{LL;K_{1A}}_{aK}\sin\theta_{K_1}+M^{LL;K_{1B}}_{aK}\cos\theta_{K_1})(\xi_uC_2-\xi_t(C_4+C_{10})) \non  &&
-\xi_t(M^{SP;K_{1A}}_{aK}\sin\theta_{K_1}+M^{SP;K_{1B}}_{aK}\cos\theta_{K_1})(C_6+C_8)\non  &&+f_B
(F^{LL;K_{1A}}_{aK}\sin\theta_{K_1}+F^{LL;K_{1B}}_{aK}\cos\theta_{K_1})(\xi_ua_1-\xi_t(a_3+a_5+a_7+a_{9}))  \non &&
-\xi_tf_B(F^{LL;K_{1A}}_{aK}\sin\theta_{K_1}+F^{LL;K_{1B}}_{aK}\cos\theta_{K_1})(a_3+a_5-\frac{1}{2}a_7-\frac{1}{2}a_9)\non  &&
-\xi_t(M^{LL;K}_{aK_{1A}}\sin\theta_{K_1}+M^{LL;K}_{aK_{1B}}\cos\theta_{K_1})(C_4-\frac{1}{2}C_{10})\non  &&
-\xi_t(M^{SP;K}_{aK_{1A}}\sin\theta_{K_1}+M^{SP;K}_{aK_{1B}}\cos\theta_{K_1})(C_6-\frac{1}{2}C_8).\label{kk3}
\en
In Eq.(\ref{kk3}), if exchanging the positions of
$K_{1A}(K_{1B})$ and $K$, we will get the total amplitude of the decay $\bar B^0\to K_1(1270)^{-} K^+$.
The total amplitudes of the decays $B\to K_1(1400)K$ can be obtained by making the replacements with
$\sin\theta_{K_1}\to\cos\theta_{K_1}, \cos\theta_{K_1}\to -\sin\theta_{K_1}$ in Eqs.(\ref{kk1}-\ref{kk3}), respectively.
\end{appendix}

\end{document}